\documentclass[showpacs,preprint]{revtex4-1}
\usepackage{hyperref}
\usepackage{amsfonts}
\usepackage{amsmath}
\usepackage{amssymb}
\usepackage{graphicx}

\begin{document}

\title{Deriving the gravitational field equation and horizon entropy for
arbitrary diffeomorphism-invariant gravity from spacetime solid}
\author{Shao-Feng Wu$^{1}$\footnote{%
Corresponding author. Email: sfwu@shu.edu.cn; Phone: +86-021-66136202.}, Bin
Wang$^{2}$, Xian-Hui Ge$^{1}$, and Guo-Hong Yang$^{1}$}
\affiliation{$^{1}$ Department of Physics, Shanghai University, Shanghai 200444, China\\
$^{2}$ Department of Physics, Fudan University, Shanghai 200433, China}

\begin{abstract}
Motivated by the analogy between the spacetime and the solid with
inhomogeneous elasticity modulus, we present an alternative method to obtain
the field equation of any diffeomorphism-invariant gravity, by extremizing
the constructed entropy function of the displacement vector field of
spacetime solid. In general stationary spacetimes, we show that the Wald
entropy of horizon arises from the on-shell entropy function of spacetime
solid.
\end{abstract}

\maketitle

\section{Introduction}

There has been growing interest recently in studying the relation between
gravitational field equations describing bulk spacetime dynamics and horizon
thermodynamics. A pioneer work on this topic was done in \cite{Jacobson}
where Einstein's equation emerges as an equation of state from the basic
thermodynamic relation in the Rindler spacetime. A lot of effort has been
spent to quantify this relation including studies in the so called $f\mathbf{%
(R)}$ gravity \cite{Eling} and the scalar tensor gravity \cite{Akbar}.
Moreover, it has been found that the relation between gravity and
thermodynamics exists also in other spacetimes, including a general static
spherically symmetric spacetime \cite{Pad0}, the dynamical Vaidya spacetime
\cite{Cai2} and cosmological spacetime \cite{Frolov,Cai,Gong,Wu}, where more
modified gravity theories like Lovelock gravity \cite{Pad0} and braneworld
gravity \cite{Cao1,Sheykhi1,Ge} have been described by the first law of
thermodynamics on the black hole horizon or cosmological apparent horizon.
Recently, it was further found that for arbitrary diffeomorphism-invariant
gravity theories, the field equation can be obtained as a state equation of
Rindler horizon thermodynamics \cite{Brustein,Parikh,Pad4}. The disclosed
relation between horizon thermodynamics and bulk gravitational field
equation can shed the light on holography \cite{Wang} and even may change
the understanding of gravity. In fact, the puzzling thermodynamic feature of
gravity and/or spacetime is one important motivation of the proposal that
gravity might not be a fundamental interaction but rather an emergent large
scale/numbers phenomenon \cite{Dreyer}.

In thermodynamics, the equations governing the equilibrium state of the
system can be obtained by the extremization of the entropy function. In
terms of the connection between gravity and thermodynamics, it seems natural
that in the gravity context one could also define an entropy function(al)
for spacetime and derive the gravitational field equation from it to
describe the spacetime. A novel approach to realize this idea was presented
in \cite{Pad1,Pad2}, where they made an analogy between the spacetime and an
elastic solid. The idea of spacetime-solid analogy is not new but has a long
history, since Sakharov's induced gravity, see \cite{Volovik}. At the
microscopic level of the analogous picture, its associated degrees of
freedom, the `spacetime atoms', are still elusive, but its existence is
crucial to interpret the thermal phenomenon associated with spacetime. At
the macroscopic level, what may be really interesting is that, one can
expect to develop a theory of elasticity phenomenologically to describe the
spacetime solid, envisaging gravity as analogous to the elasticity.
Obviously, the fundamental dynamical variable of spacetime solid is the
displacement vector field which describes the elastic deformation of the
solid, while not the metric, which may be thought of as a coarse grained
description of the spacetime at macroscopic scales, somewhat like the
density of a solid. Although the analogy is not strong, Padmanabhan et al.
\cite{Pad1,Pad2} successfully constructed an entropy function of a
displacement vector field and obtained the correct field equations of
Einstein and Lovelock gravity theories by a special variational principle on
the entropy function. Moreover, when evaluated on-shell for a solution
admitting a horizon, the extreme value of the entropy agrees with standard
Wald entropy \cite{Wald,Wald2}. This justifies that the constructed elastic
entropy has the significance of gravity entropy. Beside further supporting
the deep connection between gravity and thermodynamics, this approach is
attractive since it has the following two ingredients: (a) the metric is not
the dynamical variable; (b) the field equations remain invariant under the
shift $T_{ab}\rightarrow T_{ab}+\bar{\lambda}g_{ab}$ of the matter
energy-momentum tensor $T_{ab}$. It was pointed out that these ingredients
may be important to solve the cosmological constant problem \cite{Pad2}.

The spacetime-solid analogy formalism proposed in \cite{Pad1,Pad2} leads to
the gravitational field equations restricted in Einstein theory and Lovelock
theory. The standard Wald entropy in \cite{Pad1} is obtained only on the
local Rindler frame. Considering that the thermodynamic derivation of field
equation can be applicable to arbitrary diffeomorphism-invariant gravity
theories, it is natural to ask: Does the proposed formalism hold in all
spacetimes beyond Einstein gravity and Lovelock theory? Does it imply
something in deep? In this work we are going to address these questions. We
will analogize the spacetime as an inhomogeneous elastic solid, and develop
a general formalism to obtain the theory for arbitrary
diffeomorphism-invariant gravity. Taking the homogeneous limit, our
formalism can reduce to obtain the Einstein and Lovelock gravity theories
got in \cite{Pad1}.

The organization of the paper is the following. In section II, we define the
entropy function of a displacement vector field in the spacetime solid for
arbitrary diffeomorphism-invariant gravity. Then in section III, we
extremize the entropy function to derive the equilibrium equation. In
section IV, we compute the extreme of entropy function on general static and
non-static spacetimes and show that under appropriate circumstances it is
identical to the expression for the Wald horizon entropy. We conclude our
results in the last section.

\section{Entropy function for any diffeomorphism-invariant gravity}

Now we will describe the spacetime as a deformed solid at the macroscopic
level. Usually, to describe a general deformable solid with the theory of
elasticity, one can define the thermodynamical function, like the entropy
(or free energy, internal energy etc.) with the displacement vector field $%
\xi ^{a}\left( x\right) $ which describes the elastic displacement of the
solid through the equation $x^{a}\rightarrow x^{a}+\xi ^{a}\left( x\right) $%
, to capture the relevant dynamics in the long-wavelength limit. Varying the
thermodynamical function with respect to $\xi ^{a}$, one can obtain the
equilibrium equation of $\xi ^{a}$ (see the standard book of elastic
mechanics by Landau and Lifshitz \cite{Landau}). Analogizing spacetime to
solid, the equilibrium equation of spacetime solid can be expected. However,
what is remarkable is that the field equations will appear from the
equilibrium equation \cite{Pad1}. Before achieving this, the key task is how
to specify the thermodynamical function of spacetime.

There are several restrictions for the form of thermodynamical function. At
first, the thermodynamical function of the spacetime solid should be a
scalar, preserving the covariance of equilibrium equation. Second, in the
theory of elasticity, the thermodynamical function can be written as an
integral over a quadratic function of small strain tensor, preserving the
function density to be translationally invariant. In \cite{Pad1} the entropy
function of spacetime solid was proposed as%
\begin{equation}
S_{g}\left[ \xi \right] \sim \int_{V}d^{D}x\sqrt{-g}\left( P^{abcd}\nabla
_{c}\xi _{b}\nabla _{d}\xi _{b}\right) ,  \label{Sg}
\end{equation}%
by treating $\nabla _{a}\xi _{b}$ as the strain tensor and $P^{abcd}$ as the
elasticity modulus. However, in the presence of non-gravitational matter
distribution in spacetime, one can not demand the translational invariance
of entropy density. Hence, the entropy density can have quadratic terms in
both the derivatives $\nabla _{a}\xi _{b}$ as well as $\xi _{b}$ itself. So
it was proposed that the total entropy function should also include the
matter entropy function \cite{Pad1}%
\begin{equation*}
S_{m}\left[ \xi \right] \sim \int_{V}d^{D}x\sqrt{-g}T^{ab}\xi _{a}\xi _{b}.
\end{equation*}%
Assuming the \textquotedblleft constancy\textquotedblright\ conditions for
elasticity modulus $\nabla _{d}P^{abcd}=0$ analogous to a homogeneous (and
isotropy) solid, it was argued that the total entropy function can be used
to describe the Einstein and Lovelock gravity \cite{Pad1}.

In our study here we are going to release the constraint $\nabla
_{d}P^{abcd}=0$, which is analogous to treat the spacetime as an
inhomogeneous solid. By this nontrivial extension we will achieve to obtain
gravitational field equations for more general gravity theories. Before
constructing the concrete form of entropy function, it is natural to expect
to get some hints from the inhomogeneous elastic theory. In mechanics of
materials, the single and homogeneous materials can be fabricated as
piezocomposite materials. The interface between materials produces an uneven
distribution of stresses which reduces the electric-field-induced
displacement characteristics. It is interesting to see that these materials,
the called Functionally Graded Materials \cite{FGM}, have a total free
energy:%
\begin{equation}
F=\int_{V}d^{3}x\left( \frac{1}{2}P^{ijkl}\varepsilon _{ij}\varepsilon
_{kl}-\varepsilon ^{ijk}\varepsilon _{ij}E_{k}-\frac{1}{2}%
K^{ij}E_{i}E_{j}\right) ,
\end{equation}%
where $\varepsilon _{ij}$ denotes the components of the strain tensor, $E_{i}
$ is the component of the electric field vector related with the
displacement vector by the constitutive equations. The quantities $P^{ijkl}$
and $\varepsilon _{ijk}$ represent the elastic and the piezoelectric
modulus, respectively, and $K^{ij}$ is the electric permittivity. Motivated
by this free energy, we present that a general entropy function of spacetime
solid should be
\begin{equation}
S_{g}\left[ \xi \right] \sim \int_{V}d^{D}x\sqrt{-g}\left( P^{abcd}\nabla
_{c}\xi _{a}\nabla _{d}\xi _{b}+\nabla _{d}P^{abcd}\nabla _{c}\xi _{a}\xi
_{b}+\nabla _{c}\nabla _{d}P^{acbd}\xi _{a}\xi _{b}\right) .  \label{S0}
\end{equation}%
Thus roughly one may make the analogy of the last two terms in (\ref{S0}) as
the piezoelectric effect of inhomogeneous spacetime solid.

We need to determine the elasticity modulus $P^{abcd}$, especially its
symmetry which, in the usual elastic solid, will characterize symmetry of
structures of crystals, such as monoclinic and tetragonal, etc. However, the
concrete form of $P^{abcd}$ only can be determined in a complete theory by
the long wavelength limit of the microscopic theory just as the elastic
constants can in principle be determined from the microscopic theory of the
lattice. In macroscopic level, one can only know that the structure of the
gravitational sector is encoded in the form of $P^{abcd}$, so the object $%
P^{abcd}$ should be built out of metric and other geometric quantities. Such
a tensor can be constructed as a series in the powers of the derivatives of
the metric. One may expect that the lowest order term leads to Einstein's
theory while higher order terms come from the quantum corrections of
underlying microscopic theory. In \cite{Pad1}, assuming $P^{abcd}$ has
symmetry with Riemann tensor $R^{abcd}$ and impose $\nabla _{d}P^{abcd}=0$,
the lowest and second order correspond to Einstein's theory and Gauss-Bonnet
theory (then Lovelock theory), respectively. However, if we release the
condition $\nabla _{d}P^{abcd}=0$, more general tensor are possible. The
general construction of $P^{abcd}$ should be universal for arbitrary
diffeomorphism-invariant gravity, including the tensor for Einstein gravity
and Lovelock gravity as our special cases. Assuming the symmetry of $%
P^{abcd} $ in identical with $R^{abcd}$, a simplest candidate is%
\begin{equation}
P^{abcd}=\frac{\partial L}{\partial R_{abcd}},  \label{P4}
\end{equation}%
where $L$ is the Lagrangian of gravity theories.

Finally, the total entropy can be written as\
\begin{eqnarray}
S[\xi ] &=&S_{g}\left[ \xi \right] +S_{m}\left[ \xi \right]   \label{S} \\
&=&4\int_{V}d^{D}x\sqrt{-g}\left( P^{abcd}\nabla _{c}\xi _{a}\nabla _{d}\xi
_{b}+\nabla _{d}P^{abcd}\nabla _{c}\xi _{a}\xi _{b}+\nabla _{c}\nabla
_{d}P^{acbd}\xi _{a}\xi _{b}-\frac{1}{4}T^{ab}\xi _{a}\xi _{b}\right) ,
\notag
\end{eqnarray}%
where some proportional constants are chosen with hindsight. It should be
pointed out that the entropy function Eq. (\ref{Sg}) for Einstein and
Lovelock gravity and their generalization Eq. (\ref{S}) for general gravity
can not be fixed completely by the analogy with elastic theory. One can
understand that the entropy functions Eqs. (\ref{Sg}) and (\ref{S}) are
constructed phenomenologically. One of the desired phenomenon is to obtain
the field equation from the equilibrium equation. The other is to justify
the entropy function with the significance of gravity entropy, which is
implemented by identifying the on-shell entropy function with the Wald
horizon entropy. We will show both of them in sections below.

\section{Field equations from extremizing the entropy}

In this section, we will derive the equilibrium equation of spacetime solid.
Although the expression (\ref{S}) is well defined for any displacement
vector field, one can only obtain significant results for suitably chosen
vector fields. The vector is required to characterize the special property
of spacetime solid. The most nontrivial property of spacetime is the
existence of the horizons which act as one-way membranes which block
information for a specific class of observers. The existence of horizons,
which are null hypersurfaces, is a feature of any geometrical theory of
gravity and is reasonably independent of the field equations. We hence
assume that the spacetime solid is deformed induced by the change of
horizon. As the simplest case, one can consider that the matter is freely
falling into the horizon along the transverse invariant ingoing geodesics.
Alternatively, one can also consider a virtual displacement of horizon
radially normal to itself engulfing the matter. Obviously in this case the
total displacement of spacetime solid is induced by the change of horizon.
On the true horizon, this change, i.e. the displacement vector $\xi _{a}$,
should be characterized by its outward null normal vectors. Actually, we
will derive the correct field equations when we impose $\xi _{a}$ as null
vectors and obtain the entropy of horizon when we use the outward unit
normal of a near horizon surface to approach the null normal of true horizon.

Varying the vector field $\xi _{a}$ after adding a Lagrangian multiplier $%
\lambda (x)$ for imposing $\xi _{a}$ with constant null norm $\delta \left(
\xi _{a}\xi ^{a}\right) =0$, we find:%
\begin{eqnarray*}
\delta S[\xi ] &=&4\int_{V}d^{D}x\sqrt{-g}[2P^{abcd}\nabla _{c}\xi
_{a}\nabla _{d}\delta \xi _{b}+\nabla _{d}P^{abcd}\nabla _{c}\xi _{a}\delta
\xi _{b}+\nabla _{d}P^{abcd}\nabla _{c}\delta \xi _{a}\xi _{b} \\
&&+\nabla _{c}\nabla _{d}P^{acbd}(\delta \xi _{a}\xi _{b}+\xi _{a}\delta \xi
_{b})-\frac{1}{2}\left( T^{ab}+\lambda g^{ab}\right) \xi _{a}\delta \xi _{b}]
\\
&=&4\int_{V}d^{D}x\sqrt{-g}[2\nabla _{d}(P^{abcd}\nabla _{c}\xi _{a}\delta
\xi _{b})-2P^{abcd}\nabla _{d}\nabla _{c}\xi _{a}\delta \xi _{b}-2\nabla
_{d}P^{abcd}\nabla _{c}\xi _{a}\delta \xi _{b}+\nabla _{d}P^{abcd}\nabla
_{c}\xi _{a}\delta \xi _{b} \\
&&+\nabla _{c}(\nabla _{d}P^{abcd}\delta \xi _{a}\xi _{b})-\nabla _{c}\nabla
_{d}P^{abcd}\delta \xi _{a}\xi _{b}-\nabla _{d}P^{abcd}\delta \xi _{a}\nabla
_{c}\xi _{b}-\nabla _{c}\nabla _{d}\left( 2P^{acdb}+P^{abcd}\right) \xi
_{b}\delta \xi _{a} \\
&&-\frac{1}{2}\left( T^{ab}+\lambda g^{ab}\right) \xi _{a}\delta \xi _{b}].
\end{eqnarray*}%
where we have used the symmetry of Riemann tensor $P^{abcd}=P^{[ab][cd]}$, $%
P^{abcd}=P^{cdab}$. And we get the second to the last term above following%
\begin{equation}
\nabla _{c}\nabla _{d}P^{acbd}(\delta \xi _{a}\xi _{b}+\xi _{a}\delta \xi
_{b})=-\nabla _{c}\nabla _{d}\left( P^{cabd}+P^{cbad}\right) \xi _{a}\delta
\xi _{b}=-\nabla _{c}\nabla _{d}\left( 2P^{acdb}+P^{abcd}\right) \xi
_{a}\delta \xi _{b},  \label{index}
\end{equation}%
by using rotational symmetry of $P^{a[bcd]}=0$. Then $\delta S[\xi ]$ can be
simplified as
\begin{eqnarray*}
\delta S[\xi ] &=&4\int_{V}d^{D}x\sqrt{-g}[2\nabla _{d}(P^{abcd}\nabla
_{c}\xi _{a}\delta \xi _{b})+\nabla _{c}(\nabla _{d}P^{abcd}\delta \xi
_{a}\xi _{b})-2P^{abcd}\nabla _{d}\nabla _{c}\xi _{a}\delta \xi _{b} \\
&&-2\nabla _{c}\nabla _{d}P^{acdb}\xi _{a}\delta \xi _{b}-\frac{1}{2}\left(
T^{ab}+\lambda g^{ab}\right) \xi _{a}\delta \xi _{b}],
\end{eqnarray*}%
which leads, under the stokes theorem,%
\begin{eqnarray*}
\delta S[\xi ] &=&4\int_{\partial V}d^{D-1}x\sqrt{h}[2n_{d}(P^{abcd}\nabla
_{c}\xi _{a}\delta \xi _{b})+n_{c}(\nabla _{d}P^{abcd}\delta \xi _{a}\xi
_{b})] \\
&&-8\int_{V}d^{D}x\sqrt{-g}[P^{abcd}\nabla _{d}\nabla _{c}\xi _{a}\delta \xi
_{b}+\nabla _{c}\nabla _{d}P^{acdb}\xi _{a}\delta \xi _{b}+\frac{1}{4}\left(
T^{ab}+\lambda g^{ab}\right) \xi _{a}\delta \xi _{b}],
\end{eqnarray*}%
where $n_{a}$ is the vector outward normal to boundary $\partial V$ and $h$
is the determinant of the intrinsic metric on $\partial V$. As usual, we set
the variation $\delta \xi _{a}$ to zero at boundary (Even though the $\xi
_{a}$ is not fixed on the boundary, we still do not care about the boundary
term since it only gives the boundary condition of $\xi _{a}$, which will
not affect the bulk equilibrium equation which is independent of $\xi _{a}$
as we will show.). Therefore the first integration in the above equation
vanishes, and the condition that $S[\xi ]$ be an extremum for arbitrary
variations of $\xi _{a}$ leads%
\begin{eqnarray}
\delta S[\xi ] &=&-8\int_{V}d^{D}x\sqrt{-g}\left[ P^{abcd}\nabla _{d}\nabla
_{c}\xi _{a}\delta \xi _{b}+\nabla _{c}\nabla _{d}P^{acdb}\xi _{a}\delta \xi
_{b}+\frac{1}{4}\left( T^{ab}+\lambda g^{ab}\right) \xi _{a}\delta \xi _{b}%
\right]   \notag \\
&=&4\int_{V}d^{D}x\sqrt{-g}\left[ P^{bedc}R_{\;edc}^{a}-2\nabla _{c}\nabla
_{d}P^{acdb}-\frac{1}{2}\left( T^{ab}+\lambda g^{ab}\right) \right] \xi
_{a}\delta \xi _{b},  \label{detS}
\end{eqnarray}%
where we have used the definition of the Riemann tensor in terms of
commutator of covariant derivatives, as well as some alteration of index.
One can find that the equilibrium equation that follows from our variation
principle of entropy function is%
\begin{equation}
\left( P^{bedc}R_{\;edc}^{a}-2\nabla _{c}\nabla _{d}P^{acdb}-\frac{1}{2}%
T^{ab}-\frac{1}{2}\lambda g^{ab}\right) \xi _{a}=0.  \label{eqm1}
\end{equation}%
Requiring the condition (\ref{eqm1}) to hold for arbitrary null vector field
$\xi _{b}$ (It is interesting to note that this requirement is also invoked
in the derivation of field equation as a state equation in \cite{Jacobson}),
one finds that the equilibrium equation from extremizing the total entropy
is reduced to%
\begin{equation}
P^{bedc}R_{\;edc}^{a}-2\nabla _{c}\nabla _{d}P^{acdb}-\frac{1}{2}T^{ab}-%
\frac{1}{2}\lambda g^{ab}=0.  \label{eqm2}
\end{equation}%
This is a remarkable result that we have obtained a dynamical equation
governing the background instead of the usual equilibrium equation determine
the displacement vector field. This uninstinctive situation happens because
the symmetry of tensor $P^{abcd}$ and the entropy function are so special
that Eq. (\ref{eqm1}) does not contain derivatives with respect to $\xi ^{a}$%
.

By demanding conservation of the stress tensor and using the Bianchi
identities, one can find that the Lagrangian multiplier%
\begin{equation}
\lambda =L-\frac{\Lambda }{8\pi G},  \label{Lamb}
\end{equation}%
where $\Lambda $ is an integration constant and $G$ is the Newton
gravitational constant. One can immediately see that the equation (\ref{eqm2}%
) is just the exact field equation derived from ordinary variational
principle from the action of arbitrary diffeomorphism invariant theories of
gravity \cite{Wald2} (involving no more than the second derivatives of the
spacetime metric $g_{ab}$ and the first derivatives of the matter fields $%
\Psi _{m}$)%
\begin{equation*}
I=\int_{V}d^{D}x\sqrt{-g}L(g_{ab},R_{abcd},\Psi _{m},\nabla _{a}\Psi _{m}),
\end{equation*}%
supplemented by appropriate generalizations of Gibbons-Hawking-like boundary
terms. One can find two crucial features of our derivation that, the
variational principle is based on a vector with constant norm in spacetime
instead of the usual metric field $g_{ab}$, and the equilibrium equation is
invariant under the shift $T_{ab}\rightarrow T_{ab}+\bar{\lambda}g_{ab}$
since $\bar{\lambda}g_{ab}\xi ^{a}\xi ^{b}=\bar{\lambda}\varepsilon $ is not
varied when $\xi ^{a}$ is varied, regardless of whether $\nabla
_{d}P^{abcd}=0$.

\section{On-shell entropy function}

The result in the previous section provides an alternative variational
principle in deriving field equations of arbitrary diffeomorphism invariant
theories of gravity. In the following, we will show that the boundary term
of entropy function $S\left[ \xi \right] $ will lead to the standard Wald
entropy. The specific case on this topic was discussed in \cite{Pad1}.

Manipulating the covariant derivatives of Eq. (\ref{S}), the entropy
function can be rewritten as%
\begin{eqnarray}
S\left[ \xi \right] &=&4\int_{V}d^{D}x\sqrt{-g}(P^{abcd}\nabla _{c}\xi
_{a}\nabla _{d}\xi _{b}+\nabla _{d}P^{abcd}\nabla _{c}\xi _{a}\xi
_{b}+\nabla _{c}\nabla _{d}P^{acbd}\xi _{a}\xi _{b}-\frac{1}{4}T^{ab}\xi
_{a}\xi _{b})  \notag \\
&=&4\int_{V}d^{D}x\sqrt{-g}\nabla _{d}\left( P^{abcd}\nabla _{c}\xi _{a}\xi
_{b}\right) -4\int_{V}d^{D}x\sqrt{-g}(P^{abcd}\nabla _{d}\nabla _{c}\xi
_{a}\xi _{b}+\nabla _{c}\nabla _{d}P^{acbd}\xi _{a}\xi _{b}-\frac{1}{4}%
T^{ab}\xi _{a}\xi _{b})  \notag \\
&=&4\int_{\partial V}d^{D-1}xn_{d}\left( P^{abcd}\nabla _{c}\xi _{a}\xi
_{b}\right) +2\int_{V}d^{D}x\sqrt{-g}(P^{bedc}R_{\;edc}^{a}-2\nabla
_{c}\nabla _{d}P^{acdb}-\frac{1}{2}T^{ab})\xi _{b}\xi _{a}  \notag \\
&=&4\int_{\partial V}d^{D-1}xn_{d}\left( P^{abcd}\nabla _{c}\xi _{a}\xi
_{b}\right) +\int_{V}d^{D}x\sqrt{-g}\lambda \varepsilon ,  \label{S1}
\end{eqnarray}%
where similar derivations of Eq. (\ref{index}) has been used in the second
equality, and the field equation (\ref{eqm2}) has been used in getting the
last line. This result shows that, as mentioned in \cite{Pad4}, the second
term of third line can be thought of as the entropy in any
diffeomorphism-invariant gravity theory, except a total divergence. Then one
can recover the entropy function Eq. (\ref{S}) by reversing the derivation
in Eq. (\ref{S1}).

One can find that the final result of Eq. (\ref{S1}) is not affected by $%
\nabla _{d}P^{abcd}\neq 0$. The second term of the last line vanishes since $%
\varepsilon =0$ (Even if $\varepsilon $ is a nonvanishing constant, it is
not a surface term \cite{Pad1,Pad2}). Therefore, we will concentrate on the
first term, which will be interpreted as the surface entropy of horizon.

At this stage, we have not put any restriction of the boundary $\partial V$.
In general, the boundary is $\left( D-1\right) $-dimensional. Hence the
entropy function should not be the desired entropy of some $\left(
D-2\right) $-dimensional section of horizons. However, it was pointed out in
\cite{Pad1} that, when part of the boundary $\partial V$ is null, it is
intrinsically $\left( D-2\right) $-dimensional. This case needs to be
handled by a limiting procedure from near horizon to true horizon.

In \cite{Pad1}, the standard Wald entropy of Einstein and Lovelock gravities
are obtained from the boundary term on the local Rindler frame through a
limiting process to the null Rindler horizon. They also mentioned that the
same result can be recovered for any static spherically symmetric spacetime.
Here we will extend their discussion to the general static spacetime without
spherical symmetry and even to the stationary but non-static spacetime.

We will briefly introduce the coordinate system which is suited for the
discussion of the general static spacetime (see the details in \cite{Medved}%
). Given the general static spacetime, one can decompose the metric into a
block-diagonal form as follows%
\begin{equation*}
ds^{2}=-N^{2}dt^{2}+g_{\mu \nu }dx^{\mu }dx^{\nu },\;\mu ,\nu =1,2,\cdots
\end{equation*}%
We can arbitrarily choose a particular $\left( D-2\right) $-surface in the
constant-time slice and utilize Gaussian normal coordinates in the
surrounding region,%
\begin{equation*}
g_{\mu \nu }dx^{\mu }dx^{\nu }=dn^{2}+g_{AB}dy^{A}dy^{B},\;A,B=2,3,\cdots
\end{equation*}%
where $n$ represents the spatial direction normal to the specified $\left(
D-2\right) $-surface. The Killing horizon $H$, generated by the timelike
Killing vector field $\chi =\partial _{t}$ is approached as $%
N^{2}\rightarrow 0$. One can verify that $\kappa \equiv \left. \partial
_{n}N\right\vert _{n\rightarrow 0}$ complies with the standard version of
the surface gravity. This enables us to write a near horizon Taylor
expansion for the lapse $N(n,y)=\kappa n+O(n^{2})$.\ Since we want the
horizon to be regular and not possess a curvature singularity, some
curvature invariants must remain finite in the horizon limit, which enables
us to refine the expansion for the lapse as%
\begin{equation}
N(n,y)=\kappa n+\frac{\kappa _{2}(y)}{3!}n^{3}+O(n^{4}),  \label{N}
\end{equation}%
and write%
\begin{equation}
g_{AB}(n,y)=\left[ g_{H}\right] _{AB}(y)+\frac{\left[ g_{2}\right] _{AB}(y)}{%
2!}n^{2}+O(n^{3}).  \label{gab}
\end{equation}%
One can find that the metric (\ref{N}) and (\ref{gab}) are more general than
the metric of local Rindler frame in the presence of $\kappa _{2}$ and $%
\left[ g_{2}\right] _{ab}$. The later directly leads to the nonvanishing
component of $\nabla _{b}\xi _{d}$, not only $\nabla _{t}\xi _{t}$, which
makes the identification in \cite{Pad1} become invalid between the entropy
function and Wald entropy through introducing binormal to the cross section
of $H$.

We shall evaluate the surface integral for $S\left[ \xi \right] _{on-shell}$
near Killing horizon $H$ on a timelike surface $\Sigma $, which is denoted
as $n=$constant and the neighboring spacetime is described by metric (\ref{N}%
) and (\ref{gab}). This timelike surface $\Sigma $ can be called as the
stretched horizon \cite{Thorne,Parikh1}, which has a non-singular induced
metric and then provides a more tractable boundary than the true horizon. A
rigorous one-to-one correspondence between points on the true and stretched
horizons can be realized by using, for example, the null rays that pierce
both surfaces. Roughly, we can expect that the on-shell entropy will match
the Wald entropy under the limit $n\rightarrow 0$ in the end of the
calculation. Take $\xi ^{a}=n^{a}$ as the spacelike normal to these surfaces
$\Sigma $, with components%
\begin{equation}
\xi ^{a}=n^{a}=(0,1,0,0,\ldots )  \label{zn}
\end{equation}%
and unit norm. In the limiting process the spacelike unit vector will be a
null vector, which denotes that we are considering the null surface as a
limit of a sequence of timelike surfaces. The metric determinant $h$ of
these surfaces $\Sigma $ can be decomposed as $\sqrt{h}=N\sqrt{\sigma }$,
where $\sigma $ is the metric determinant on the transverse spatial
surfaces, having the limit on the true horizon $\sqrt{\sigma }\rightarrow
\sqrt{g_{H}}$.

Now we will check whether the surface term%
\begin{equation}
S_{Pad}=4\int_{\Sigma }d^{D-1}x\sqrt{h}n_{d}\left( P^{abcd}\nabla _{c}\xi
_{a}\xi _{b}\right)  \label{PadS}
\end{equation}%
will be reduced to Wald entropy in the limiting process. At first, we will
introduce the Wald entropy \cite{Wald,Wald2} based on a simplified version
of the formalism \cite{Cardoso}. Consider a generally covariant Lagrangian $%
L $, that involves no more than quadratic derivatives of the spacetime
metric $g_{ab}$. Under the diffeomorphism $x^{a}$ $\rightarrow x^{a}+\chi
^{a}$ the metric changes via $\delta g_{ab}=-\nabla _{a}\chi _{b}-\nabla
_{a}\chi _{b}$. By diffeomorphism-invariance, the change in the action, when
evaluated on-shell, is given only by a surface term. This leads to a
conservation law, $\nabla _{a}J^{a}=0$, for which we can write $J^{a}$ $=$ $%
\nabla _{b}J^{ab}$. Here $J^{ab}$ defines (not uniquely) the antisymmetric
Noether potential associated with the diffeomorphism $\chi ^{a}$, which can
be formulated as%
\begin{equation*}
J^{ab}=-32\pi \left( P^{abcd}\nabla _{c}\chi _{d}-2\chi _{d}\nabla
_{c}P^{abcd}\right) .
\end{equation*}%
Associated with a rigid diffeomorphism $\chi ^{a}$, there is the Noether
charge defined by integrating the Noether potential over any closed
spacelike surface $S$ of codimension two. It turns out that the
corresponding Noether charge is just proportional to the entropy%
\begin{equation}
S_{Wald}=\frac{1}{8\kappa }\int_{B}J^{ab}dB_{ab},  \label{WaldS}
\end{equation}%
when $\chi ^{a}$ is a timelike Killing vector (with vanishing norm at the
Killing horizon), and the surface $B$ is the cross-section of Killing
horizon $H$. However one can formally define the quantity $S_{Wald}$ on any
closed spacelike surface $\Omega $, and only in the end take the limit in
which $\Omega $ approaches a section $B$ of the Killing horizon $H$. In the
following, we will define such a quantity on the section $\Omega $ of a
stretched horizon $\Sigma $ described by metric (\ref{N}) and (\ref{gab}),
and compare $S_{Pad}$ with $S_{Wald}$ in the limit $n\rightarrow 0$. On the
stretched horizon $\Sigma $, the timelike Killing vector can be specified as%
\begin{equation}
\chi ^{a}=(1,0,0,0,\ldots ).  \label{K}
\end{equation}%
The proper velocity $u^{a}$ of a fiducial observer moving along the orbit of
$\chi ^{a}$ is $u^{a}=\chi ^{a}/\sqrt{-g_{ab}\chi ^{a}\chi ^{b}}=\left(
\frac{d}{d\tau }\right) ^{a}$ where $\tau $ is the proper time. The fiducial
proper velocity $u^{a}$ and unit normal $n^{a}$ of stretched horizon $\Sigma
$ define $d\Omega _{ab}=n_{(a}u_{b)}dA$, where $dA=\sqrt{\sigma }d^{D-2}y$
is the area element on cross section $\Omega $. We will evaluate both $%
S_{Pad}$ and $S_{Wald}$ for several typical diffeomorphism invariant
theories of gravity. For simplicity, we will restrict on $D=4$, but the
results can be directly generalized to more higher dimensions.

\subsection{$L\sim f(\protect\phi ,R)$}

As the first example let us consider the following Lagrangian
\begin{equation*}
L=\frac{1}{16\pi G}f(\phi ,R).
\end{equation*}%
Obviously, this example contains the popular $f(R)$ and scalar-tensor
gravity as its special cases. From the Lagrangian the tensor $P^{abcd}$
reads as%
\begin{equation}
P^{abcd}=\frac{\partial L}{\partial R_{abcd}}=\frac{1}{32\pi G}\frac{%
\partial f}{\partial R}\left( g^{ac}g^{bd}-g^{ad}g^{bc}\right) .  \label{P1}
\end{equation}%
Substituting this tensor and the normal vector Eq. (\ref{zn}) into Eq. (\ref%
{PadS}), and preserving the leading term of $n$ in the end of the
calculation, we obtain%
\begin{eqnarray*}
S_{Pad} &=&4\int_{\Sigma }d^{3}x\sqrt{h}n_{d}\left( P^{abcd}\nabla
_{c}n_{a}n_{b}\right) \\
&=&\int_{\Sigma }d^{3}x\sqrt{g_{H}}\frac{\partial f}{\partial R}\left[ \frac{%
\kappa }{8\pi G}+O\left( n^{2}\right) \right] .
\end{eqnarray*}%
Restricting the $t$ integral within the range $(0,2\pi /\kappa )$ for the
periodicity in Euclidean time \cite{Hawking,Astefanesei}, we can obtain%
\begin{equation*}
S_{Pad}\simeq \int_{\Omega }\frac{1}{4G}\frac{\partial f}{\partial R}\sqrt{%
g_{H}}d^{2}y.
\end{equation*}%
When $f(\phi ,R)=R$, the entropy density $\frac{1}{4\pi G}\frac{\partial f}{%
\partial R}$ is $\frac{1}{4\pi G}$, and $S_{Pad}$ is the well-known
Gibbons-Hawking entropy. Substituting Eq. (\ref{P1}), Killing vector (\ref{K}%
), and $d\Omega _{ab}=n_{(a}u_{b)}dA$ into Wald entropy (\ref{WaldS}), and
preserving the leading term of $n$ at last, we get%
\begin{eqnarray}
S_{Wald} &=&\frac{1}{8\kappa }\int_{\Omega }J^{ab}d\Omega _{ab}  \label{S3}
\\
&=&\int_{\Omega }\frac{\partial f}{\partial R}\left[ \frac{1}{4G}+O\left(
n^{2}\right) \right] \sqrt{g_{H}}d^{2}y.  \notag
\end{eqnarray}%
We notice that although the higher-order terms $O\left( n^{2}\right) $ of $%
S_{Pad}$ and $S_{Wald}$ are different, their leading terms are exactly the
same.

\subsection{$L\sim \protect\alpha R_{abcd}R^{abcd}+\protect\beta %
R_{ab}R^{ab} $}

Beside the term $f(R)\sim R^{2}$, the higher (quadratic) derivative
interactions usually include%
\begin{equation*}
L=\frac{1}{16\pi G}\left( \alpha R_{abcd}R^{abcd}+\beta R_{ab}R^{ab}\right) ,
\end{equation*}%
with arbitrary parameters $\alpha $, $\beta $, which may depend on some
scalar fields. Its derivative with respect to $R_{abcd}$ is%
\begin{equation*}
P^{abcd}=\frac{1}{8\pi G}\left[ \alpha R^{abcd}+\frac{1}{4}\beta \left(
g^{bd}R^{ac}-g^{ad}R^{bc}+g^{ac}R^{bd}-g^{bc}R^{ad}\right) \right] .
\end{equation*}%
Similar to the above case, we can obtain%
\begin{eqnarray}
S_{Pad} &=&\int_{\Omega }[\frac{-\beta }{2Gg_{H}}(\left[ g_{2}\right] _{22}%
\left[ g_{H}\right] _{33}+\left[ g_{H}\right] _{22}\left[ g_{2}\right] _{33}+%
\frac{2\kappa _{2}}{\kappa }\left[ g_{H}\right] _{22}\left[ g_{H}\right]
_{33}-2\left[ g_{H}\right] _{23}\left[ g_{2}\right] _{23}-\frac{2\kappa _{2}%
}{\kappa }\left[ g_{H}\right] _{23}^{2})  \label{S2} \\
&&-\frac{2\alpha \kappa _{2}}{G\kappa }+O\left( n^{2}\right) ]\sqrt{g_{H}}%
d^{2}y,  \notag
\end{eqnarray}%
which is identical with the evaluation of $S_{Wald}$ up to $O\left(
n^{2}\right) $.

In \cite{Pad1}, it was proved that the Wald entropy of Gauss-Bonnet gravity $%
L\sim R_{abcd}R^{abcd}-4R_{ab}R^{ab}+R^{2}$ is identical with the entropy
derived from Eq. (\ref{PadS}) (in Rindler frame). We have checked that this
is a special case of combination of our discussions in A and B, since the
formula $S_{Wald}$ and $S_{Pad}$ are both linear in the Lagrangian.

\subsection{$L\sim R_{ab}\protect\nabla ^{a}\protect\phi \protect\nabla ^{b}%
\protect\phi $}

It is known that there are some ambiguities in Wald entropy. Considering the
following interaction involving the metric and a scalar field%
\begin{equation*}
L=\frac{1}{16\pi G}R_{ab}\nabla ^{a}\phi \nabla ^{b}\phi ,
\end{equation*}%
the corresponding $P^{abcd}$ reads%
\begin{equation*}
P^{abcd}=-\frac{1}{16\pi G}\nabla ^{\lbrack a}\phi g^{b][c}\nabla ^{d]}\phi .
\end{equation*}%
It is important to notice that this tensor is not altered if one adds some
terms about scalar fields (but no more than their two order derivative) into
$L$, such as,
\begin{equation*}
L_{i}=\frac{1}{16\pi G}\left[ \nabla ^{a}\nabla ^{b}\phi \nabla _{a}\nabla
_{b}\phi -\left( \nabla ^{2}\phi \right) ^{2}+R_{ab}\nabla ^{a}\phi \nabla
^{b}\phi \right] .
\end{equation*}%
In \cite{Jacobson93}, it was pointed out that, $L_{i}\epsilon $ ($\epsilon $
is the volume form) can be written as a total derivative $L_{i}\epsilon
=d\alpha _{i}$, which yields a vanishing contribution to the Wald entropy.
This contradicts with the direct evaluation using the tensor $P^{abcd}$ in
Wald entropy on the section of Killing horizon, until one realizes that $%
L_{\chi }\phi =0$ on it. For the entropy $S_{Pad}$, one can evaluate%
\begin{eqnarray}
S_{Pad} &=&-\frac{1}{4\pi G}\int_{\Sigma }d^{3}x\sqrt{h}n_{d}\left( \nabla
^{\lbrack a}\phi g^{b][c}\nabla ^{d]}\phi \nabla _{c}n_{a}n_{b}\right)
\notag \\
&=&\int_{\Omega }\left[ \frac{\partial _{n}^{2}\phi }{8G}+O\left( n\right) %
\right] \sqrt{g_{H}}d^{2}y,  \label{S4}
\end{eqnarray}%
which is identical with the evaluation of $S_{Wald}$ up to $O\left( n\right)
$, and seems nonvanishing. In fact, one can make the similar understanding
to that in \cite{Jacobson93}, where $S_{Pad}\sim n_{a}\nabla ^{a}\phi $
should vanish, since $n_{a}$ will be a null Killing vector when $%
n\rightarrow 0$, and hence $n_{a}\nabla ^{a}\phi \sim L_{\chi }\phi $
vanishes on the Killing horizon.

Hereto, we have shown that the surface term $S_{Pad}$ can be reduced to Wald
entropy near the static horizon in the leading order (the higher order terms
are different). We will further show that this result can be generalized to
any stationary but not static black holes. For such a black hole, it is
expected to be axially symmetric \cite{Waldbook}. Consider our spacetime is
invariant under \textquotedblleft time-reversal\textquotedblright , it is
convenient to write the spacetime metric near horizon $n=0$ as \cite{Medved1}
\begin{equation*}
ds^{2}=-N\left( n,z\right) ^{2}dt^{2}+g_{\phi \phi }\left( n,z\right) \left[
d\varphi -\omega \left( n,z\right) dt\right] ^{2}+dn^{2}+g_{zz}dz^{2},
\end{equation*}%
where $N$ denotes the usual lapse\ function and $\omega $ is the
angular-rotation\ parameter. The zeroth law of black hole mechanics and
rigidity theorem \cite{Carter} for axially symmetric (stationary,
non-static) Killing horizons tell that surface gravity $\kappa \equiv \left.
\partial _{n}N\right\vert _{n\rightarrow 0}$ and $\omega _{H}\equiv \left.
\omega \right\vert _{n\rightarrow 0}$ must be a non-negative constant on the
horizon, respectively. Moreover, a stationary Killing horizon is a geodesic
submanifold which implies that the horizon is extrinsically flat, i.e. the
extrinsic curvature of the horizon must be zero $\left. K_{\mu \nu
}\right\vert _{n\rightarrow 0}=0$. The above properties directly imply a set
of necessary constraints%
\begin{eqnarray}
g_{\varphi \varphi }(n,z) &=&\left[ g_{H}\right] _{\varphi \varphi }(z)+%
\frac{\left[ g_{2}\right] _{\varphi \varphi }(z)}{2!}n^{2}+O(n^{3}),  \notag
\\
g_{zz}(n,z) &=&\left[ g_{H}\right] _{zz}(z)+\frac{\left[ g_{2}\right]
_{zz}(z)}{2!}n^{2}+O(n^{3}),  \notag \\
\omega (n,z) &=&\omega _{H}+\frac{\omega _{2}(z)}{2!}n^{2}+O(n^{3}),  \notag
\\
N(n,z) &=&\kappa n+\frac{\kappa _{2}(z)}{3!}n^{3}+O(n^{4}),  \label{g1}
\end{eqnarray}%
where the first-order terms in $\omega $ and $N$ are required to vanish to
avoid a curvature singularity on the horizon. The Killing vector of this
spacetime is%
\begin{equation}
\chi ^{a}=(1,\omega _{H},0,0),  \label{K1}
\end{equation}%
in terms of $\left( t,\varphi ,n,z\right) $ coordinate system.

Invoking the time integration carried out in Euclidean sector, and using the
metric (\ref{g1}), the Killing vector (\ref{K1}) and the corresponding
fiducial velocity $u^{a}$, we can evaluate the surface term $S_{Pad}$ and
Wald entropy near the horizon for different gravity theories. We find the
entropy on horizon:%
\begin{equation*}
S_{Pad}=\int_{\Omega }\frac{\partial f}{\partial R}\left[ \frac{1}{4G}%
+O\left( n^{2}\right) \right] \sqrt{g_{H}}dzd\varphi =S_{Wald}+O\left(
n^{2}\right) ,\text{ for case A,}
\end{equation*}%
\begin{eqnarray*}
S_{Pad} &=&\int_{\Omega }[\frac{-\beta }{4Gg_{H}}(\left[ g_{2}\right]
_{\varphi \varphi }\left[ g_{H}\right] _{zz}-\frac{1}{\kappa ^{2}}\left[
g_{H}\right] _{\varphi \varphi }^{2}\left[ g_{H}\right] _{zz}\omega
_{2}(z)^{2}+\left[ g_{H}\right] _{\varphi \varphi }\left[ g_{2}\right] _{zz}+%
\frac{\kappa _{2}}{\kappa }\left[ g_{H}\right] _{\varphi \varphi }\left[
g_{H}\right] _{zz}) \\
&&+\frac{\alpha }{4G\kappa ^{2}}(3\left[ g_{H}\right] _{\varphi \varphi
}^{2}\omega _{2}-4\kappa \kappa _{2})+O\left( n^{2}\right) ]\sqrt{g_{H}}%
dzd\varphi \\
&=&S_{Wald}+O\left( n^{2}\right) ,\text{ for case B,}
\end{eqnarray*}%
\begin{equation*}
S_{Pad}=\int_{\Omega }\left[ \frac{\partial _{n}^{2}\phi }{8G}+O\left(
n\right) \right] \sqrt{g_{H}}dzd\varphi =S_{Wald}+O\left( n\right) ,\text{
for case C.}
\end{equation*}%
Now we can conclude that the boundary term $S_{Pad}$ are the same as the
standard Wald entropy for general static and stationary but non-static black
holes. This is one of the key results of our paper, which justifies our
function $S\left[ \xi \right] $ as an authentic `entropy'.

\section{Conclusion and discussion}

In this paper, we have generalized the analogy between the spacetime and
solid developed in \cite{Pad1}. We have shown that the spacetime with
generalized gravity theory can be analogized to the inhomogeneous elastic
solid. Spacetimes described by the Einstein and Lovelock gravity are the
limiting cases when we consider the solid being homogeneous. One key point
in this work is that we have successfully constructed an entropy function of
a displacement vector field to describe phenomenologically the macroscopic
level of spacetime solid. Extremizing the total entropy function with
respect to the displacement vector, we have found that the equilibrium
equation can be identified with the field equation of general
diffeomorphism-invariant gravity. Our generalization supports that there is
an analogy between spacetime and solid and that the gravitational
thermodynamics is not restricted on the concrete gravity theory \cite%
{Brustein}. We also expect that this approach is helpful to obtain the field
equation in the scenarios of emergent gravity where the correct field
equation is absent \cite{Dreyer}.

Beside providing an alternative approach to obtain the arbitrary
gravitational field equation, we have shown that the entropy function on the
boundary of any stationary spacetime is identical with the standard Wald
horizon entropy in several typical higher derivative theories of gravity,
though a general relationship between these two entropy expressions still
needs further understanding. This provides a new method to calculate the
horizon entropy for any stationary spacetime besides the Hamiltonian method,
Noether charge method and the field redefinition method \cite{Jacobson93}.

It should be stressed that the spacetime solid is very special. This solid
acquires the equilibrium equation independent of the displacement vector
field. We suspect that this acquirement may be related to certain underlying
first principle to constrain the macroscopic description of spacetime solid.
Moreover, the displacement vector field has the zero norm, which
characterizes the key role taken by the null hypersurfaces in the spacetime
solid \cite{Frittelli,Cremades}. Furthermore, the spacetime solid has a
special surface, the stretched horizon. Technically, we have only treated it
as the near horizon to approach the true horizon. However, in the membrane
paradigm, it has been treated as a dynamic fluid membrane, obeying such
pre-relativistic equations as Ohm's law and the Navier-Stokes equation \cite%
{Thorne,Parikh1}. Obviously, it is valuable to study the thermodynamical
property of stretched horizon in the spacetime solid, comparing the
differences and similarities with the fluid membrane. If one can understand
these nontrivial properties of spacetime solid, the analogy might be of use
in the context of the semiclassical limit and even the quantum gravity
(spacetime) enigma. In fact, recently it have been proposed \cite{Pad3} that
if the entropy function for Einstein and Lovelock gravity is interpreted as
an action in the semiclassical limit, its value will affect the phase of the
semiclassical wave function, then the observer independent of the
semiclassical gravity requires this phase (i.e. the Wald horizon entropy) to
be quantized in units of $2\pi $. Our work further suggests that the Wald
entropy can be similarly quantized in arbitrary diffeomorphism-invariant
theories of gravity.

\begin{acknowledgments}
The work of BW was supported by the NSFC. The work of SFW and XHG were
partially supported by NSFC under Grant Nos. 10905037 and 10947116. The work
of GHY, XHG and SFW was partially supported by Shanghai Leading Academic
Discipline Project (project number S30105) and Shanghai Research Foundation
No 07dz22020.
\end{acknowledgments}

\end{document}